\documentclass[fleqn,10pt]{wlscirep}
\title{Quantum man-in-the-middle attack on the calibration process of quantum key distribution}

\author[1]{Yang-Yang Fei}
\author[1]{Xiang-Dong Meng}
\author[1,*]{Ming Gao}
\author[1]{Hong Wang}
\author[1,2]{Zhi Ma}
\affil[1]{State Key Laboratory of Mathematical Engineering and Advanced Computing, Zhengzhou, Henan, 450001, China}
\affil[2]{CAS Center for Excellence and Synergetic Innovation Center in Quantum Information and Quantum Physics, University of Science and Technology of China, Hefei, Anhui, 230026, China}

\affil[*]{gaoming@meac-skl.cn}

\begin{abstract}
Quantum key distribution (QKD) protocol has been proved to provide unconditionally secure key between two remote legitimate users in theory. Key distribution signals are transmitted in a quantum channel which is established by the calibration process to meet the requirement of high count rate and low error rate. All QKD security proofs implicitly assume that the quantum channel has been established securely. 
 However, the eavesdropper may attack the calibration process to break the security assumption of QKD and provide precondition to steal information about the final key successfully. 
Inspired by N. Jain et al., Phys. Rev. Lett.107,110501(2011), we reveal the security risk of the calibration process of a passive-basis-choice BB84 QKD system by launching a quantum man-in-the-middle attack which intercepts all calibration signals and resends faked ones. Large temporal bit-dependent or basis-dependent detector efficiency mismatch can be induced.  Then we propose a basis-dependent detector efficiency mismatch (BEM) based faked states attack on a single photon BB84 QKD to stress the threat of BEM. Moreover, the security of single photon QKD systems with BEM is studied simply and intuitively. Two effective countermeasures are suggested to remove the general security risk of the calibration process.
\end{abstract}
\begin{document}

\flushbottom
\maketitle
%
%
\thispagestyle{empty}

\section*{Introduction}

Comparing with traditional communication protocols, quantum key distribution (QKD) protocol has been proven to have unconditional security to distribute key between two remote parties, known as Alice and Bob, with the assumption that all devices are perfect\cite{gisin2002quantum,lutkenhaus2009focus,scarani2009security}. However, imperfections, i.e., the differences between the theoretical model and the practical implementation, widely exist in practical QKD systems due to technological limitations and improper protocol operations, which can be exploited by the eavesdropper, Eve, to break the overall security of QKD.\par
Nowadays, many kinds of quantum hacking strategies are proposed exploiting different imperfections, which include photon number splitting attack\cite{huttner1995quantum,sun2011proof}, Trojan horse attack\cite{gisin2006trojan, jain2014trojan}, phase-remapping attack\cite{fung2007phase, xu2010experimental}, partially random phase attack\cite{sun2012partially,tang2013source,sun2014hacking} on the source and detector-efficiency-mismatch based attack\cite{makarov2005faked, makarov2006effects, makarov2008faked, qi2007time, zhao2008quantum}, detector control attack \cite{lydersen2010hacking,lydersen2010thermal,sauge2011controlling,makarov2009controlling,lydersen2011controlling}, side channel attack\cite{lamas2007breaking} and others\cite{wiechers2011after, weier2011quantum, li2011,sun2011passive, wang2013effect}. The results help to improve the security of practical QKD systems and make them more reliable. Nevertheless, most quantum hackings focus on the key distribution process and ignore other important aspects of QKD security, such as the establishment of the indispensable quantum channel, except one\cite{jain2011device}.\par
All QKD security proofs assume the quantum channel has been established securely. In fact, the establishment of the quantum channel, especially the calibration of devices in long distance communication, can be hacked by Eve. Although no key information is transmitted in the calibration process, Eve may attack this process to induce security loophole and provide precondition to mount quantum hacking successfully in the following QKD process. So the security of practical QKD systems is closely related to the calibration process, which is needed to be studied to assure no loophole is induced before key distribution. Normally, legitimate users do not check the legitimacy of the calibration signals because only statistical property can be monitored. Thus, quantum man-in-the-middle attack on the improper design of device calibration is always hard to be discovered by legitimate users.\par
Almost all QKD systems need to calibrate the activation timing of gated-mode detectors before key distribution. The activation timing of multiple detectors differs slightly due to temperature related drift in electronic chips with limited precision and discrepancies in the lengths of fibers connecting them. Ideally, the differences of the activation timing of multiple detectors should take constant values to minimize the detector efficiency mismatch \cite{zhao2008quantum}. However, the lengths of optical fibers and the values set in electronic chips vary with time and temperature\cite{stucki2011long}, which makes immobilization of the differences almost impossible. So QKD systems with multiple detectors always scan the activation timing of gated-mode detectors independently to reduce the detector efficiency mismatch occurred naturally in time frame\cite{zhao2008quantum,jain2011device,dixon2010continuous}. This design works well without Eve and a large detector efficiency mismatch occurs with the probability of only 4${\%}$\cite{qi2007time}. However, this convenient design may also be utilized by Eve to induce large detector efficiency mismatch in the calibration process.\par
In Ref. [\citen{jain2011device}], by hacking the calibration process of a phase coding "plug \& play" system, named as Clavis2 \cite{Clavis2}, a temporal separation of 450 ps of the detection efficiencies is induced. However, because of the short pulse width of the calibration light, the separation between detectors is still limited. Moreover, the security of the calibration process of passive-basis-choice BB84 QKD systems has not been investigated. In this paper, inspired by N. Jain et al.\cite{jain2011device}, we propose a quantum man-in-the-middle attack strategy on the activation timing calibration process and experimentally reveal the security risk of the independent scanned activation timing on a passive-basis-choice BB84 QKD system by separating the efficiency curves of detectors in two bases, i.e., basis dependent detector efficiency mismatch (BEM). Then a BEM-based faked states attack (FSA) on single photon BB84 QKD is also proposed. Finally, we analyze the security of single photon QKD systems with BEM simply and intuitively. The corresponding secure key rate formula is also given out.\par
Measurement device independent (MDI) QKD \cite{lo2012measurement} can remove all the side channels in the detection part, which can also make our attack on the calibration process useless. However, MDI QKD has high requirement for equipment, which makes MDI QKD protocol not as practical as traditional decoy-state BB84 protocol. Thus most practical (commercial) QKD systems still use BB84 protocol with decoy states\cite{wang2005beating,lo2005decoy,hwang2003quantum}. Therefore, studying the security of the calibration process of practical QKD systems is still of great importance.\par

\section*{Quantum man-in-the-middle attack on the calibration process}

The establishment of quantum channel is indispensable before key exchange in all QKD systems with gated-mode single photon detectors. To satisfy the count rates requirement of legitimate users, the detectors are needed to be calibrated in time dimension via a process, i.e., the line length measurement (LLM) in Ref. [\citen{jain2011device}], to align the activation timing of the detectors with the arrival timing of the signals. In normal LLM process, Alice (Bob) sends calibration signals to estimate the channel length and relative delay between the arrival timing of the pulses at different detectors. The activation timing of each detector is scanned electronically and independently (while monitoring count rates) to find out the timing when the count rates are maximum. The quantum channel is also under Eve's control in the LLM process and legitimate users do not check the legitimacy of the calibration signals since these signals contain no information.\par
\begin{figure}[!htbp]
  \centering
  \includegraphics[width=0.8\textwidth]{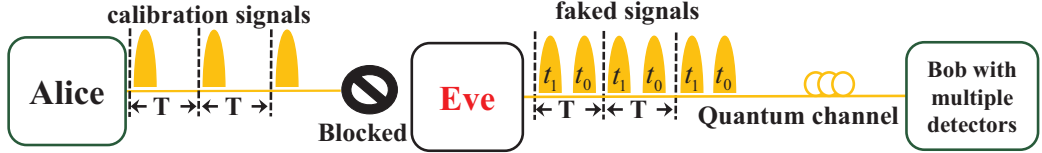}
  \caption{(Color online). Simple diagram of our quantum man-in-the-middle attack strategy on the calibration process. T: one cycle.}\label{fig:1}
\end{figure}\par
 Usually, these calibration signals only contain one pulse in each cycle in QKD systems (in phase encoding QKD systems the calibration signals is one pair in each cycle, which can also be treated as one). So what if Eve induces some disparities between the activation timing of different detectors by replacing the calibration signals with faked ones which contain more than one pulses in each cycle? This is the origin of our attack. \par
In our quantum man-in-the-middle attack on the LLM process, Eve blocks all the calibration signals and sends faked calibration signals to disturb the activation timing calibration of detectors. As shown in Fig. \ref{fig:1}, the calibration signals sent by Alice only contain one pulse in each cycle while the faked signals sent by Eve have two (the responding arrival timing are denoted as $t_0$ and $t_1$).
Eve will not be discovered by legitimate users as long as the count rates are compensated by simply adjusting the average photon number of the faked calibration signals. \par
In theory, the scanned "activation-timing v.s. efficiency" curves of detectors in the LLM process without Eve have only one peak. However, the "activation-timing v.s. efficiency" curves of detectors under faked signals have two similar peaks (the corresponding timing is $t_0$ and $t_1$). Considering the fluctuations of detector efficiencies and statistical fluctuation due to finite signals, the maximum count rate occurs at $t_0$ or $t_1$ randomly. So the activation timing of each detector is also set at $t_0$ or $t_1$ randomly.  \par
In a passive-basis-choice BB84 QKD system, two kinds of efficiency mismatch can be induced, which includes BEM and detector efficiency mismatch in the same basis (DEM). So there are four situations after our attack. For simplicity, let $p(t_x=t_y)$, $x$${\in}$${\{H,V,+,-\}}$, $y$${\in}$${\{0,1\}}$, represent the probability that the activation timing of detector $x$ is set at $t_y$ and $p(t_x=t_y)$=1/2 in theory. The first situation is no detector efficiency mismatch, i.e., all the activation timing of four detectors are set at the same timing ($t_0$ or $t_1$), and the corresponding probability is $p(t_H=t_0)p(t_V=t_0)p(t_+=t_0)p(t_-=t_0)+p(t_H=t_1)p(t_V=t_1)p(t_+=t_1)p(t_-=t_1)$=$(1/2)^4+(1/2)^4$=1/8. The second one is partial DEM with a probability of 1/2, which means that DEM exists in only one basis. The third one is that DEM exists in both bases, and the probability is $p(t_H=t_0)p(t_V=t_1)p(t_+=t_0)p(t_-=t_1)+p(t_H=t_0)p(t_V=t_1)p(t_+=t_1)p(t_-=t_0)$+$p(t_H=t_1)p(t_V=t_0)p(t_+=t_1)p(t_-=t_0)+p(t_H=t_1)p(t_V=t_0)p(t_+=t_0)p(t_-=t_1)$=1/4. The last situation is BEM, of which the probability of $p(t_H=t_0)p(t_V=t_0)p(t_+=t_1)p(t_-=t_1)+p(t_H=t_1)p(t_V=t_1)p(t_+=t_0)p(t_-=t_0)$=$(1/2)^4+(1/2)^4$=1/8. All the situations can be utilized by Eve except the first one. By this trick, Eve is able to induce large temporal efficiency mismatch in the LLM process when the activation timing of different detectors are set at different values. If the interval between $t_0$ and $t_1$ is large enough, we may totally separate the efficiency curves of two detectors. \par
For the sake of attacking QKD process successfully, Eve need to know the activation timing of each detector ($t_0$ or $t_1$). This can be solved by time shift attack (TSA)\cite{qi2005timeshift,qi2007time} in the beginning of the key exchange process, in which Eve randomly shifts the arrival timing of signals in key-distribution process to $t_0$ or $t_1$. Note that one advantage of TSA is that it introduces no additional errors because Eve does not impact the signals \cite{qi2005timeshift,qi2007time}. However, TSA will reduce the overall detection rate. In the extreme case where the efficiency curves of two bases are completely separated, the detection rate of each detector reduces to 1/2 of the original one. But in practice, Eve can compensate the decrease of the detection rate by changing the transmission of the channel. Thus, Eve may stay hidden during key exchange. In QKD process, Alice and Bob compare the "basis" value of each click over the classical channel in the $sifting$ step and reveal a random sample of the bits of their raw keys to estimate the error rate in the quantum channel in the $error$ $rate$ $estimation$ step. So Eve knows some clicks in each detector as well as the arrival timing of the pulses shifted by herself. If one detector always clicks when the arrival timing is shifted to $t_0$ ($t_1$) and seldom clicks when the arrival timing is shifted to $t_1$ ($t_0$), Eve can confirm that the activation timing of this detector is set to $t_0$ ($t_1$). Therefore, Eve obtains the actual setting of the activation timing of each detector.\par
 After confirming the activation timing of each detector, Eve can use different methods to steal information about the final key in the rest key exchange process. If the efficiency mismatch induced is confirmed to be DEM, Eve can obtain information about the raw key by mounting TSA and analyzing the arrival timing of each pulse shifted by herself. However, if the efficiency mismatch induced is confirmed to be BEM (in a passive-basis-choice system), which can not be hacked by TSA, she can also mount FSA to break the overall security of practical QKD systems.\par
\begin{figure}[htbp]
  \centering
  \includegraphics[width=0.8\textwidth]{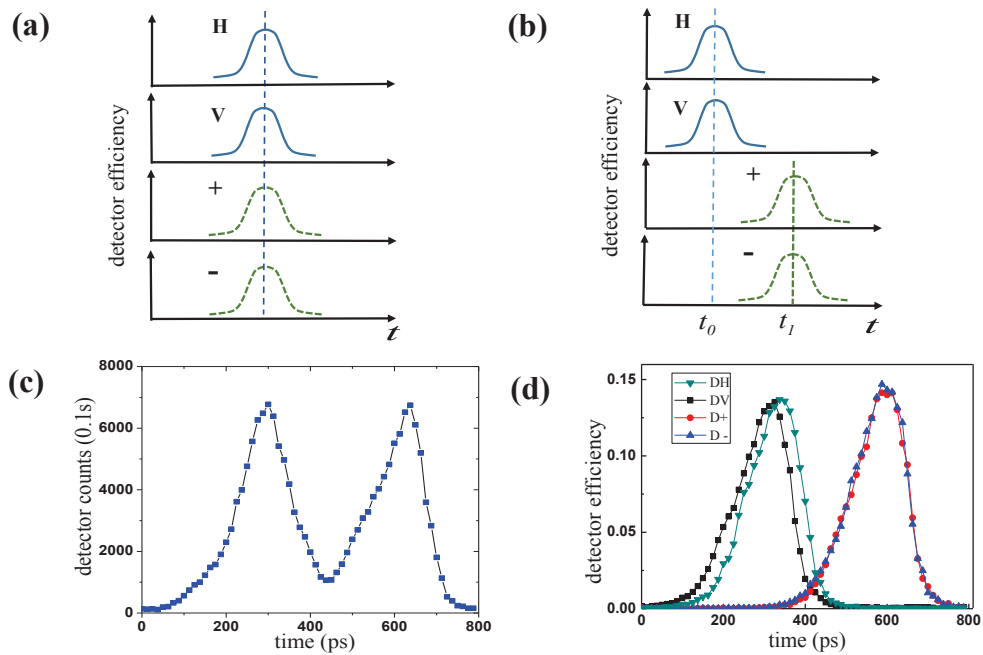}
  \caption{(Color online). (a)Sketch of efficiency curves of four detectors in time frame without being attacked after the calibration process. (b)Sketch of BEM, which is the goal of our proof-of-principle experiment. (c)The faked detector efficiency curve of one of four detectors in our quantum man-in-the-middle attack, the faked efficiency curve of detector V is given as an example. (d)BEM induced after the attack. The downward triangle marker line is the efficiency of detector H; The square marker line is the efficiency of detector V; The circle marker line is the efficiency of detector +; And the upward triangle line is the efficiency of detector -.}\label{fig:2}
\end{figure}

Security of DEM has been widely studied\cite{fung2009security,DEM2010,maroy2010security,fei2015}. Thus we mainly focus on the security of BEM. In \textbf{Method I}, we describe a proof-of-principle experiment of our quantum man-in-the-middle attack in the LLM process to induce BEM. Our goal is to separate the efficiency curves of two bases in time frame in a polarized passive-basis-choice BB84 QKD system. As shown in Fig. \ref{fig:2}(b), we take the situation that the activation timing of detectors in basis $Z$ ($X$) is $t_0$ ($t_1$) as an example.
So the overall theoretically probability of the situation as the one in Fig. \ref{fig:2}(b) is $p(t_H=t_0)p(t_V=t_0)p(t_+=t_1)p(t_-=t_1)$=$(1/2)^4$=1/16. Note that there are two kinds of BEM (the activation timing of basis $Z$ ($X$) is $t_0$ ($t_1$) and the activation timing of basis $Z$ ($X$) is $t_1$ ($t_0$)) respectively), the one shown in Fig. \ref{fig:2}(b) is one of them.


\par
A typical faked detector efficiency curve of detector V obtained in our experiment is shown in Fig.\ref{fig:2}(c), from which we can get that there are indeed two similar peaks as we expect. The interval of the two peaks is 337.5ps, which is related to the interval between ${t_0}$ and ${t_1}$. Fig. \ref{fig:2}(d) shows the actual average value of BEM induced in our proof-of-principle experiment. We repeated the experiment 320 times and got BEM as the one in Fig. \ref{fig:2}(b) 22 times, which is close to the ideal value, 20 times. \par


The experiment results show that by mounting our quantum man-in-the-middle attack on the calibration process several times, Eve is certain to induce huge BEM to benefit herself in the following QKD process. Moreover, Eve knows the exact optimum time to mount FSA ($t_0$ and $t_1$). So legitimate users should pay more attention to the calibration process of practical passive-basis-choice BB84 QKD systems (see \textbf{Conclusion and discussion} for detail).

\section*{A BEM-based FSA on single photon BB84 QKD}
We have demonstrated that BEM can be induced by Eve in the calibration process above. Here we present a BEM-based FSA on single photon BB84 QKD to stress the threat of BEM.
\subsection*{FSA strategy}
For single photon BB84 QKD systems with large BEM, Eve can mount an FSA to break the overall security to gain the information of the final key. In order to explain this strategy particularly, we suppose that the BEM, which exists in practical QKD systems, is like the one in Fig. \ref{fig:2}(b). At a place very near Alice, Eve randomly selects her basis to measure the pulses from Alice, and sends her results to Bob in the same measurement basis at different timing (denoted as $t_0$ and $t_1$). For example, if Eve measures Alice's pulse in the $Z$ ($X$) basis and her measurement result is bit 0, then she resends bit 0 in the $Z$ ($X$) basis at time $t_0$ ($t_1$). We assume the maximum efficiency mismatch occurs at $t_0$ and $t_1$ to benefit Eve below. Moreover, we grant Eve with perfect detectors and perfect single photon source. Similar with DEM-based FSA\cite{makarov2005faked}, Bob's detectors barely click when his measurement basis is not the same with Eve's. By implementing this trick, Eve is able to control Bob's measurement basis and bit.\par

\subsection*{Security of FSA on single photon QKD}
In this part, BEM-based FSA on single photon BB84 QKD systems is analyzed. To calculate the count rate and the quantum bit error rate ($QBER$, denoted as $e_b$) after the attack, we use the same method used in Ref.[\citen{jain2011device}] and generalize Table \ref{tab:1} (see {\bf{Methods II}} for detail). To simplify our calculation, dark counts of the detectors are ignored and symmetry of the efficiencies of detectors is also assumed. ${\eta_m(t_n)}$, $m$${\in}$${\{X,Z\}}$,$n$${\in}$${\{0,1\}}$, represents the equivalent overall transmission and detection efficiency between Alice and Bob's basis $m$ at time $t_n$.From Table 1, we can obtain that when Eve mounts BEM-based FSA, the count rate of Bob is
$$
{p_{arrive}}=\frac{1}{4}[\eta_Z(t_1) + \eta_Z(t_0) + \eta_X(t_1) + \eta_X(t_0)],\eqno{(1)}
$$
and the error rate is

$$
{p_{error}}=\frac{1}{8}[\eta_Z(t_1) + \eta_X(t_0)].\eqno{(2)}
$$
And the corresponding overall $QBER$ after attack is given by
$${e_b} = \frac{p_{error}}{p_{arrive}} = \frac{\eta}{2(1+\eta)},\eqno{(3)}$$
where ${\eta = {\eta_Z(t_1)}/{\eta_X(t_1)} = {\eta_X(t_0)}/{\eta_Z(t_0)}}$ represents the maximum efficiency mismatch between two bases. The FSA strategy is an intercept-resend attack, and it constitutes a Markov chain(Alice $\rightarrow$ Eve $\rightarrow$ Bob). So the mutual information between Alice and Eve, $I$(A:E), is always no less than the mutual information between Alice and Bob, $I$(A:B). In symmetric QKD protocols with one-way classical communications, Alice and Bob may expect $e_b$ to be less than 11\% when they are not aware of the existing of BEM and think that the bit and phase error rates are equal (11\% is the zero of $1-2h(e_b)$)\cite{Shor2000Simple,makarov2006effects,DEM2010}. Here we also use $e_b$ ${<}$ 11${\%}$ to estimate $\eta$ simply and compare with other analyses fairly\cite{makarov2006effects,DEM2010}(see the next section for security analysis and key formula of QKD systems with BEM). Thus, we get that the security of the QKD system is broken when ${\eta < 0.282}$. While in the DEM-based FSA on single photon QKD, the corresponding security bound is ${\eta' <}$ 0.066\cite{makarov2006effects}, which is much smaller than 0.282. Here, ${\eta' }$ stands for the maximal efficiency mismatch between two detectors in the same basis. Moreover, a optimal attack combining FSA and time shift attack (TSA) \cite{qi2007time, zhao2008quantum} on a QKD system with DEM requires ${\eta' < 0.252}$ \cite{DEM2010}, which is still smaller than 0.282. Our calculation indicates that BEM-based FSA is more threatening than DEM-based FSA because of the lower $QBER$ induced. Note that TSA no longer works in the BEM situation, thus no optimal individual attack can be carried out by combining  FSA and TSA.\par
\subsection*{Security of partial FSA on single photon QKD}
\begin{figure}[ht]
  \centering
  \includegraphics[width=0.5\textwidth, angle=0]{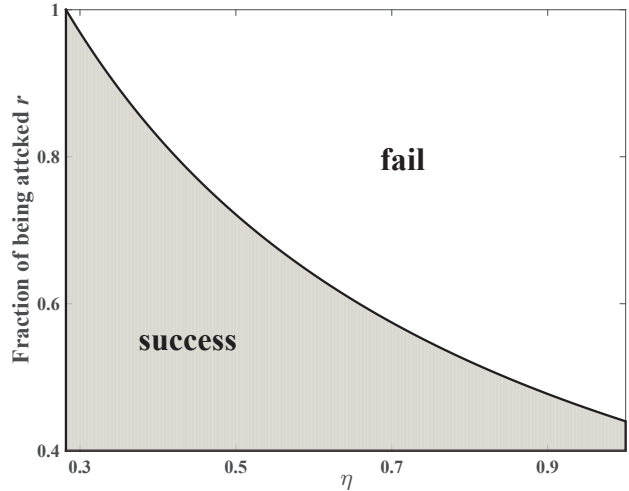}
  \caption{(Color online)The boundary of successful partial FSA.}\label{fig:new}
\end{figure}\par
\begin{figure}[ht]
  \centering
  \includegraphics[width=0.5\textwidth, angle=0]{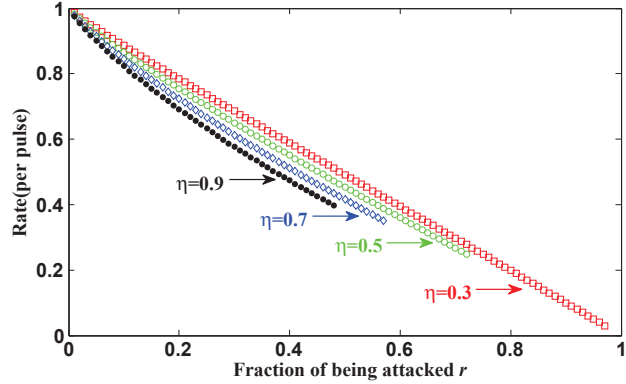}
  \caption{(Color online)The square (circle,  diamond, solid point) markers show the final key rate changes with the fraction of photons attacked by Eve when ${\eta}$=0.3 (0.5, 0.7, 0.9).}\label{fig:3}
\end{figure}\par
When ${1 \ge \eta>0.282}$, Eve mounts FSA on every photon, which make $e_b$ larger than ${11\%}$. So Eve can only attack a part of the photons and let the other photons passed without disturbing. In an one-way classical communication system, for individual attacks, according to mutual information theory, the secret key rate is given by \cite{R1,R2}
$$R = I(A:B) - I(A:E),\eqno{(4)}$$
where $I$(A:B) = 1 - $h(e_b)$, $I$(A:E) = $r$[1 - $h(q)$], $q$ is the error probability between Alice and Eve in BEM-based FSA, $r$ is the fraction attacked by Eve, ${h}(x) =  - xlo{g_2}x - (1 - x)lo{g_2}(1 - x)$ is the binary Shannon information function. From Table 1, we can get $q$ is ${\eta}/{[2(1+\eta)]}$. We also assume that the photons passing undisturbed arrive at the same time, denoted as $t_2$, and the corresponding efficiency is ${\kappa=\eta_X(t_2)=\eta_Z(t_2)=\eta_X(t_1)=\eta_Z(t_0)}$. We have
$$e_b = \frac{{r{p_{error}}}}{{(1 - r)\eta ({t_2}) + r{p_{arrive}}}} = \frac{{r\eta }}{{4 - 2r + 2r\eta }}.\eqno{(5)}$$
$e_b<11\%$ is also needed to be satisfied if Eve mounts partial FSA successfully. Fig. \ref{fig:new} shows the boundary of successful partial FSA. The shadow region shows that, for different values of $\eta$, Eve can always find proper values of $r$ and attack a part of the signal pulses without causing a detectable rise in $e_b$, i.e., she can always keep $e_b$ lower than 11\%. Fig. \ref{fig:3} shows how the final key rate changes with the fraction of photons attacked by Eve with different ${\eta}$. When ${\eta}$ gets larger, the fraction of photons can be hacked by Eve gets smaller because of the limit of $QBER$. For the same ${\eta}$, the larger fraction of photons got attacked, the less final key rate legitimate users can get.
\par

\section*{Security analysis of single photon QKD systems with BEM}
A BEM-based FSA on single photon BB84 QKD has been presented, which indicates the threat of BEM. However, the security analysis results of FSA can not be applied to other attacks. Therefore, it is very important to study the security of QKD systems with BEM and present a secure key rate formula covering BEM.\par
Security of QKD systems with DEM has been studied \cite{fung2009security,DEM2010,maroy2010security}. This kind of security analyses is based on Koashi's argument \cite{koashi1,koashi2}, which proves the security of QKD using complementarity. However, Koashi's argument requires the source and the detectors do not reveal the basis information in the BB84 protocol. Therefore, this kind of security analyses can not be adapted to the BEM situation, because BEM actually controls the probability of activating which basis in the detection part.
Other security analyses like Ref. [\citen{maroy2010security, randomness}] take the basis-dependent leakage information as a parameter related to the phase error rate. Nevertheless, it is very difficult to estimate the phase error rate $tightly$ with the basis leakage information, especially when the efficiency mismatch is large, so they are not practical in the BEM situation either (always less than 0 when efficiency mismatch is large). Besides, these analyses build different detection models, which are all difficult and abstract to understand. \par
Here, we analyze the security of single photon QKD systems with BEM.
 Indeed, BEM can be divided into two extreme cases. one case is $no$ efficiency mismatch between two bases and the secure key rate is 1-2$h(e_b)$, where $e_b$ is the overall $QBER$. The other is $complete$ efficiency mismatch when the detection efficiency of one basis is 0 and all of the information about the detection basis is revealed, so no secure key can be generated. General cases are combination of the two extreme cases, which is the key of our security analysis. So the counts of single photons can be divided into two parts, one part is from detection devices with $compelete$ efficiency mismatch and the other is from detectors with $no$ efficiency mismatch. Thus secure key rate can be extracted from the $no$ efficiency mismatch part. \par
\begin{figure}[ht]
  \centering
  \includegraphics[width=0.3\textwidth, angle=0]{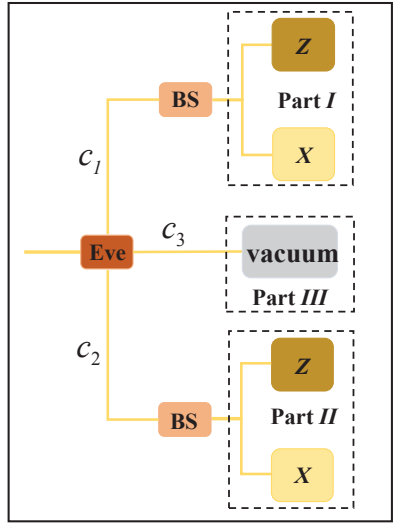}
  \caption{(Color online)The equivalent detection model at $t_0$ of single photon QKD systems. BS: beam splitter; Z: Z-basis measurement; X: X-basis measurement}\label{fig:4}
\end{figure}\par
In order to explain the method particularly, we also suppose that the BEM existing in practical QKD systems is like the one in Fig. \ref{fig:2}(b) with symmetry between bases and the single photons (after Eve's deception) arrive at ${t_0}$ or ${t_1}$ randomly to benefit Eve. We only need to analyze the secure key rate at $t_0$ because of the symmetry. To separate the counts of single photons into two parts as we stated before, an equivalent detection model of $t_0$ (shown in Fig. \ref{fig:2}(b)) is given out in Fig. \ref{fig:4}. Part $I$ corresponds the $complete$ efficiency mismatch part with $\eta_{XI}$=0 and $\eta_{ZI}$=1 and Part $II$ corresponds the $no$ efficiency mismatch part with $\eta_{XII}=\eta_{ZII}$=1, where $\eta_{jk}$, $j$${\in}$${\{X,Z\}}$,$k$${\in}$${\{I,II\}}$, represents the detection efficiency of basis $j$ in Part $k$.
Part $III$ represents vacuum measurement. Pulses transmitted into Part $III$ correspond to a part of the pulses that are not detected by Bob. 
The probability of a pulse enterinng Part $I$, $II$ and $III$ is $c_1$, $c_2$ and $c_3$ respectively and we have $c_1+c_2+c_3=1$. Note that $c_1$, $c_2$ and $c_3$ are all under Eve's control. To maximize $I(A:E)$, Eve should maintain the count rates in the equivalent detection model the same as the ones in the standard detection model, we have
$$\frac{1}{2}\eta_X(t_0)=\frac{1}{2}c_2\eta_{XII},\eqno{(6)}$$
$$\frac{1}{2}\eta_Z(t_0)=\frac{1}{2}c_1\eta_{ZI}+\frac{1}{2}c_2\eta_{ZII}.\eqno{(7)}$$
So we can get $c_1=\eta_Z(t_0)-\eta_X(t_0)$ and $c_2=\eta_X(t_0)$. Therefore, among all the clicks, the probability of one click detected in Part $I$ is $p={\frac{1}{2}c_1}/(\frac{1}{2}c_1+c_2)=(\eta_Z(t_0)-\eta_X(t_0))/(\eta_Z(t_0)+\eta_X(t_0))=(1-\eta)/(1+\eta)$. \par
We have $e_b=pe_b'+(1-p)e_b''$, where $e_b'$ ($e_b''$) is the bit error rate in Part $I$ ($II$). 
For the clicks in Part $I$, Eve can have the same measurement results as Bob, so the mutual information between Alice and Eve in Part $I$ is $1-h(e_b')$. According to the security proofs with $no$ efficiency mismatch \cite{gisin2002quantum,lutkenhaus2009focus,scarani2009security}, the maximum mutual information between Alice and Eve in Part $II$ is $h(e_p'')$, where $e_p''$ is the phase error rate in Part $II$ and $e_p''=e_b''$.  Therefore we have the overall mutual information between Alice and Eve $I(A:E)=p[1-h(e_b')]+(1-p)h(e_p'')$. To maximize $I(A:E)$, Eve lets $e_b'=0$. The overall mutual information between Alice and Bob is $I(A:B)=1-h(e_b)$ and overall maximum mutual information between Alice and Eve is $\max \limits_{Eve} I(A:E)=p+(1-p)h(e_p'')$. Finally, the secure key rate under individual attacks is given by
$$R=I(A:B)-\max\limits_{Eve}\ I(A:E)=1-h(e_b)-p-(1-p)h(e_p'')=\frac{2\eta}{1+\eta}[1-h(\frac{e_b(1+\eta)}{2\eta})]-h(e_b).\eqno{(8)}$$

We compare our analysis result with the one of Ref. [\citen{gottesman2004security}] which also studies basis-dependent flaws. The analysis can apply for imperfections on both the source and the receiver.  Eq. (50) in Ref. [\citen{gottesman2004security}] considers that the preparation bases of a fraction of photon pulses emitted by Alice are known by Eve. When applying the analysis to the receiver, the detection bases of a fraction ($\Delta$ ) of the counts are controlled by Eve. In our equivalent detection model, we give out a way to calculate $\Delta$ in the BEM situation. The detection bases of counts in Part $I$ are controlled by Eve. So $\Delta$ equals the probability that one click is detected in Part $I$ among all the clicks, i.e., $p=(1-\eta)/(1+\eta)$. And Eq. (50) in Ref. [\citen{gottesman2004security}] is changed into $R=1-\Delta-h(e_b)-(1-\Delta)h(e_b/(1-\Delta))=\frac{2\eta}{1+\eta}[1-h(\frac{e_b(1+\eta)}{2\eta})]-h(e_b)$, which is the same with Eq. (8) in our analysis.
\begin{figure}[ht]
  \centering
  \includegraphics[width=1\textwidth, angle=0]{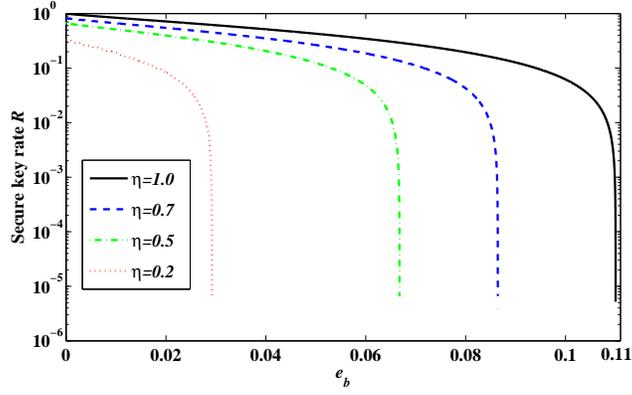}
  \caption{(Color online)The black solid (blue dash, green dash dotted, red dotted) line shows the secure key rate $R$ changes with the $e_b$ when ${\eta}$=1.0 (0.7, 0.5, 0.2).}\label{fig:5}
\end{figure}\par
Fig. \ref{fig:5} shows the secure key rate $R$ changes with $e_b$ when $\eta$ is different. When $\eta$=1, the secure key rate becomes 1-2$h(e_b)$, which is equal to the secure key rate when there is no efficiency mismatch. When the mismatch becomes larger ($\eta$ becomes smaller), less secure key can be generated. And obviously, with the same $e_b$, we will obtain less secure key rate when $\eta$ gets smaller. Moreover, when $\eta$ becomes smaller, the threshold of $e_b$ also gets smaller. Note that our analysis can also be applied to practical decoy state QKD systems with a weak coherent source by combining with the famous $GLLP$ argument\cite{gottesman2004security}.
\par
\section*{Methods}
\subsection*{I}

\begin{figure}[htbp]
  \centering
  \includegraphics[width=1.0\textwidth]{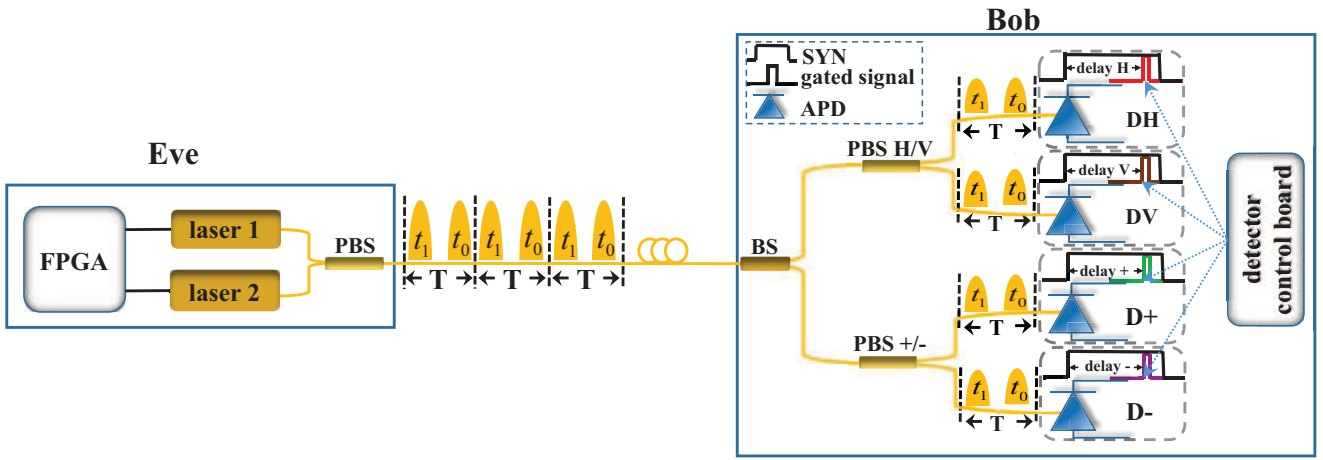}
  \caption{(Color online). Schematic diagram of our proof-of-principle experiment. FPGA: field programmable gate array; BS: beam splitter; PBS: polarization beam splitter; SYN: synchronization; P: polarizer; APD: avalanche photon detector; T: detection cycle.}\label{fig:6}
\end{figure}\par
The schematic diagram of our proof-of-principle experiment is presented in Fig. \ref{fig:6}. In polarized passive-basis-choice BB84 QKD systems, Alice sends four polarized states (H/V/+/-) randomly in the LLM process to keep the maximum count rates of four detectors equal. For simplicity, we use two lasers (laser 1 and laser 2) to produce two orthogonal polarized states with the same intensity as the faked calibration signals.  Note that we only need polarized directions of 2 lasers are orthogonal to each other. As shown in Fig. \ref{fig:6}, a polarization maintaining polarization beam splitter is used to guide pulses from lasers 1 and 2 to one optical fiber. So the polarizations of pulses from laser 1 and laser 2 are orthogonal. The wavelength of the lasers is 1550nm. Laser 1 or laser 2 randomly fires at $t_0$ and $t_1$, which is controlled by the FPGA. The detection part, Bob, can run LLM process independently. The detector control board independently scans the activation timing of each detector by changing the delay between the SYN signals and the gated signals step by step and the overall count rates at different activation timing are recorded. The value of activation timing of each InGaAs detector is set to the timing when the count rate is maximum. \par

The full width at half maximum (FWHM) of the high-speed laser is 50ps. The detection device consists of one BS, two PBSs, four single photon detectors and one detector control board. The LLM scheme works as we explained before, which is widely used in practical QKD systems\cite{zhao2008quantum,jain2011device,dixon2010continuous}. The InGaAs detectors are from $QuantumCTek$\cite{quantumctek} with a gating frequency of 1.25GHz. The scanning step of the activation timing (unit of the delays) is 12.5ps, and there are total 64 steps to cover one whole cycle. The efficiency of the detectors is about 13\% and the dark count rate is about $4\times 10^{-6}$ per gate. The frequency of our faked calibration signals is 5MHz, whose average photon number is 0.7 to compensate the count rates of detectors. And the time interval of our faked calibration signals in one cycle is about 320ps to separate the efficiency curves of two bases completely if BEM is induced.\par
\subsection*{II}
Alice sends random bit (0 or 1) in random basis ($X$ or $Z$) and Bob random measures in  $X$ or $Z$ basis. Counts are kept if Alice and Bob select the same basis. When Eve mounts BEM-based FSA on single photon QKD, we can get Table \ref{tab:1}.
\begin{table}[!htbp]
\caption{\label{tab:1}BEM-based FSA. Eve measures randomly in $X$ or $Z$ basis before mounting FSA. The first
column contains Alice's basis choice and bit value.  The second column shows Eve's measurement result. The third
column shows the parameters of the faked state resent by Eve: basis, bit, timing. The fourth column shows Bob's measurement result; The corresponding detection probabilities are shown in the last column. }
\centering  
\scriptsize
\begin{tabular}{ccccc}  
\hline
\hline
$Alice\rightarrow$ & {$\rightarrow$Eve} & {Eve$\rightarrow$} & {Bob's result} & {Detection probability}\\ \hline  
$Z$0 & {$Z$,0} & {$Z$,0,${t_0}$} & $Z$0 & ${{\eta_Z(t_0)}}$\\
&&&$Z$1&${0}$ \\

 & {$X$,0} & {$X$,0,${t_1}$} & $Z$0 & ${{\frac{1}{2}\eta_Z(t_1)}}$\\
&  &  & $Z$1 & ${{\frac{1}{2}\eta_Z(t_1)}}$\\

 & {$X$,1} & {$X$,1,${t_1}$} & $Z$0 & ${{\frac{1}{2}\eta_Z(t_1)}}$\\
&  &  & $Z$1 & ${{\frac{1}{2}\eta_Z(t_1)}}$\\

$Z$1 & {$Z$,1} & {$Z$,1,${t_0}$} & $Z$0 & ${0}$\\
&&&Z1&${{\eta_Z(t_0)}}$ \\

 & {$X$,0} & {$X$,0,${t_1}$} & $Z$0 & ${{\frac{1}{2}\eta_Z(t_1)}}$\\
&  &  & $Z$1 & ${{\frac{1}{2}\eta_Z(t_1)}}$\\

 & {$X$,1} & {$X$,1,${t_1}$} & $Z$0 & ${{\frac{1}{2}\eta_Z(t_1)}}$\\
&  &  & $Z$1 & ${\frac{1}{2}\eta_Z(t_1)}$\\

$X$0 & {$X$,0} & {$X$,0,${t_1}$} & $X$0 & ${{\eta_X(t_1)}}$\\
&&&$X$1&${0}$ \\

 & {$Z$,0} & {$Z$,0,${t_0}$} & $X$0 & ${{\frac{1}{2}\eta_X(t_0)}}$\\
&  &  & $X$1 & ${{\frac{1}{2}\eta_X(t_0)}}$\\

 & {$Z$,1} & {$Z$,1,${t_0}$} & $X$0 & ${{\frac{1}{2}\eta_X(t_0)}}$\\
&  &  & $X$1 & ${{\frac{1}{2}\eta_X(t_0)}}$\\

$X$1 & {$X$,1} & {$X$,1,${t_1}$} & $X$0 & ${0}$\\
&&&$X$1&${{\eta_X(t_1)}}$ \\

 & {$Z$,0} & {$Z$,0,${t_0}$} & $X$0 & ${{\frac{1}{2}\eta_X(t_0)}}$\\
&  &  & $X$1 & ${{\frac{1}{2}\eta_X(t_0)}}$\\

 & {$Z$,1} & {$Z$,1,${t_0}$} & $X$0 & ${{\frac{1}{2}\eta_X(t_0)}}$\\
&  &  & $X$1 & ${\frac{1}{2}\eta_X(t_0)}$\\
\hline
\hline
\end{tabular}
\end{table}

\section*{Conclusion and discussion}
In this paper, we focus on the security of the calibration process of practical QKD systems. To QKD systems with gated-mode single photon detectors, whose activation timing are scanned independently, a novel quantum man-in-the-middle attack strategy on the calibration process is proposed. By the trick of sending two faked calibration signals in one cycle, large detector efficiency mismatch can be induced. The results of proof-of-principle experiment on the detection part of a polarized passive-basis-choice QKD system show that Eve is able to totally separate the efficiencies of two bases in time frame. Moreover, we present a BEM-based FSA on single photon QKD systems to indicate the induced BEM is a serious threat to practical security of QKD. Finally, a method is proposed to analyze the security of single photon QKD systems with BEM simply and intuitively. The secure key rate formula covering BEM is given out. \par
In our BEM-based FSA, the detection basis of each pulse is controlled by Eve with a probability using timing information. Similarly, in the wavelength dependent beam splitter attack, Eve can also control the detection basis with a probability by applying different wavelengths. In particular, the blinding attack can completely control the detection bases, which corresponds to the situation that $\eta=0$ in our analysis (the secure key rate is also 0 according to our Eq. (8)). Therefore, our security analysis of QKD systems with BEM can also be adapted to wavelength dependent beam splitter attack and detector blinding attack.\par
The existing of detector efficiency mismatch seriously effects the security of QKD systems and we should pay more attention to the calibration process. To defend our quantum  man-in-the-middle attack strategy on the calibration process, the relationship of different detectors' activation timing should be strictly monitored in the calibration process. This modification is implementable in software. Besides, this kind of attacks (including our attack on the calibration process and FSA) sends faked signals to Bob at different timing. So a possible countermeasure is to monitor arrival timing of the incident pulses by using a dedicated detector \cite{makarov2006effects}. Nevertheless, this is difficult to realize and will reduce the key generation rate and increase the cost. Here, we strongly recommend a self-test method. Legitimate users install a specialized laser with an optical attenuator at Bob's side. After the calibration process, QKD system runs a self test with the specialized laser to verify whether BEM or DEM is large by sending test signals at different timing. We hope that this self-test apparatus becomes a part of future practical QKD systems.

\bibliography{sample}

\noindent 

\section*{Acknowledgements}
This work was supported by the National Natural Science Foundation of China (61472446, 61501514 and 61701539) and the Open Project Program of the State Key Laboratory of Mathematical Engineering and Advanced Computing (2016A01), the Science and Technology Projects of Henan Province(Grant No. 162102210181).

\section*{Author Contributions}
Y.-Y. F., M. G., X.-D. M. and Z. M. conceived the project. Y.-Y. F. and X.-D. M. conducted the experiment. Y.-Y. F., M. G. and H. W. performed the calculation and analysis. Y.-Y. F. wrote the paper. All authors reviewed the manuscript.

\section*{Additional Information}
\textbf{Competing financial interests:} The authors declare no competing financial interests. Correspondence and requests for materials should be addressed to M. G. (email: gaoming@meac-skl.cn).





\end{document}